\documentclass{emulateapj}

\usepackage[usenames,dvips]{color}
\usepackage[version=3]{mhchem}

\def\chandra{{\it Chandra}}
\def\sdssj01{J0100+2802}
\def\aox{$\alpha_{\text{ox}}$}

\shorttitle{}
\shortauthors{}

\begin{document}

\title{Exploratory Chandra observation of the ultraluminous quasar SDSS J010013.02+280225.8 at redshift 6.30}
\author{Yanli Ai\altaffilmark{1}, Liming Dou\altaffilmark{2}, Xiaohui Fan\altaffilmark{3,4}, Feige Wang\altaffilmark{5}, Xue-Bing Wu\altaffilmark{4,5}, Fuyan Bian\altaffilmark{6,7}}
\altaffiltext{1}{School of Physics and Astronomy, Sun Yat-Sen University, Guangzhou 510275, China, aiyanli@mail.sysu.edu.cn}
\altaffiltext{2}{Key Laboratory for Research in Galaxies and Cosmology, Astronomy Department, The University of Sciences and Technology of China, Hefei, Anhui 230026, China}
\altaffiltext{3}{Steward Observatory, University of Arizona, 933 North Cherry Avenue, Tucson, AZ 85721, USA}
\altaffiltext{4}{Kavli Institute for Astronomy and Astrophysics, Peking University, Beijing 100871, China}
\altaffiltext{5}{Department of Astronomy, School of Physics, Peking University, Beijing 100871, China}
\altaffiltext{6}{Research School of Astronomy and Astrophysics, Australian National University, Weston Creek, ACT 2611, Australia}
\altaffiltext{7}{Stromlo Fellow}


\begin{abstract}
We report exploratory \chandra\ observation of the ultraluminous quasar SDSS J010013.02+280225.8 at redshift 6.30. The quasar is clearly detected by \chandra\ with a possible component of  extended emission.
The rest-frame 2-10 keV luminosity is 9.0$^{+9.1}_{-4.5}$ $\times$ 10$^{45}$ erg s$^{-1}$ with inferred photon index of $\Gamma$ = 3.03$^{+0.78}_{-0.70}$. 
This quasar is X-ray bright, with inferred X-ray-to-optical flux ratio \aox\ $=-1.22^{+0.07}_{-0.05}$, higher than the values found in other quasars of comparable ultraviolet luminosity. 
The properties inferred from this exploratory observation indicate that this ultraluminous quasar might be growing with super-Eddington accretion and probably viewed with small inclination angle. Deep X-ray observation will help to probe the plausible extended emission and better constraint the spectral features for this ultraluminous quasar.
\end{abstract}

\keywords{quasar: individual: SDSS J010013.02+280225.8}

\section{Introduction}
A large number of quasars at redshifts of $z\ga6$ have been discovered, selected with a combination of optical and near-infrared colors \citep[][]{mortlock11,venemans13,jiang15,wang16}.
These sources are powerful probes of the cosmic reionization \citep[][]{fan06} and  offer insights into the process by which massive black holes formed and grew in early universe. 
Recently, Wu et al. (2015) reported the discovery of SDSS J010013.02+280225.8 (hereafter \sdssj01), an ultraluminous quasar with an estimated black hole mass of 12 billion solar masses. 
With the highest black hole mass and the highest luminosity among all quasars discovered at $z\gtrsim 5$, 
it sets the tightest constraints on models for the massive black hole growth and evolution at early epochs \citep[e.g.][]{shankar09, volonteri10}.

X-ray emissions from quasars reveal the conditions in the innermost regions of their accretion-disk corona, and provide information about how the black hole is fed. There are tentative evidences that high-redshift quasars have relatively soft X-ray spectra, indicating that the black holes are accreting near or above the Eddington rate \citep[][]{grupe06,page14}, although debates exist \citep[e.g.][]{just07,moretti14}.  
Mildly super-Eddington intermittent accretion may significantly ease the problem of assembling the massive black holes when the universe was young \citep[][]{madau14}. 
Furthermore, high-redshift quasars, such as the known blazars at $z>5$ \citep[][]{romani04,frey10,sbarrato13,ghisellini15}, may have powerful jet,
which provide a means by which the black hole accretes at a rate that breaks the Eddington limit \citep[][]{ghisellini13}.
The extended lobes produced by jets may be strong in X-ray, rather than in radio, as the dramatic $(1 + z)^4$ boosts in the energy density of the cosmic microwave background (CMB) causes inverse Compton scattering to dominate energy losses of relativistic electrons \citep[][]{fabian14}.

X-ray observation of \sdssj01 provides a crucial test for understanding the formation of the most massive black holes in the early universe. 
In this letter we report our exploratory \chandra\ observation of this ultraluminous quasar. 
Throughout this paper, we adopt the ${\Lambda}CDM$ cosmology parameters from 
Planck Collaboration (2014): $\Omega_{M}$ = 0.315, $\Omega_{\Lambda}$ = 0.685, and $H_{0}$ = 67.3 km s$^{-1}$.
We define power law photon index $\Gamma$ such that N(E) $\propto$ E$^{-\Gamma}$.
For the Galactic absorption of SDSS J0100+2802, which is including in the model fitting,
we use the value of  $N_{H}$ = 5.82 $\times$ $10^{20}$ cm$^{-2}$\citep[][]{kalberla05}. 
All uncertainties are given at 1$\sigma$, unless otherwise specified.

\section{Chandra observation and data analysis}
We obtained \chandra\ observation of SDSSJ0100+2802 on 2015 October 16 using the Advanced CCD Imaging Spectrometer \citep[ACIS,][]{garmire03} instrument. 
The total integration time is 14.8 ks.
The target was positioned on the ACIS-S3 chip, which was operated in 
time-exposure mode, with faint telemetry format.  The data were processed with standard CIAO version 4.7 \citep[][]{fruscione06}, and 
only events with grades of 0, 2, 3, 4, and 6 were considered in the analysis. No background flares are present in the observation.
We restrict to the derived X-ray counts in the observed-frame of 0.5-7 keV, 
considering the increasingly uncertain  quantum efficiency of ACIS
at lower energies and the steeply increasing background at higher energies. 

\begin{figure}
\includegraphics[width=0.5\textwidth]{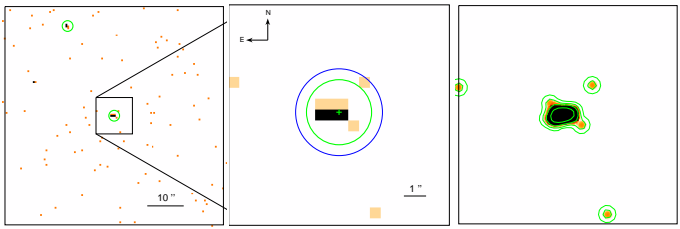}
\caption{Detection of X-ray emission of \sdssj01. Left:
A 1$\arcmin \times$1$\arcmin$ \chandra\ image centred on \sdssj01 in 0.5-7 keV. 
The circles show the two sources detected by CIAO task WAVDETECT;
Middle: the central 10$\arcsec \times$ 10 $\arcsec$ 
of the image. The plus sign shows the optical position given by SDSS. 
14 counts are detected in the 3 pixel 
radius aperture (green), and one more count in the 4 pixel radius aperture (blue). Right: the rebinned image in the middle panel to 0.1 ACIS pixel
and smoothed with a 0.492$\arcsec$ Gaussion filter.
The size of the ACIS CCD pixel is $\sim0.492\arcsec$.
\label{fig1}}
\end{figure}

We first carried out source detection with CIAO task WAVDETECT using wavelet transforms (with wavelet scale sizes of 1, 2, 4, 8
and 16 pixels) and a false-positive probability threshold of 10$^{-6}$.
The quasar is detected at equatorial coordinates of 01:00:13.038, +28:02:25.75. 
The X-ray position differs from its optical position by 0.1 arcsec, within the radius (0.6 arcsec) of 90\% uncertainty circle of \chandra\ ACIS absolute position\footnote{http://cxc.cfa.harvard.edu/cal/ASPECT/celmon/}. 
At a distance of 28$\arcsec$ to the northeast of \sdssj01, 
an X-ray source is detected with 7.9$\pm$2.8 net counts (Figure\,\ref{fig1}). 
This source is classified as a galaxy in the SDSS DR12 catalog with photometric redshift of 0.52$\pm$0.13 \citep[][]{alam15}, therefore it is likely to be a low luminosity AGN in the foreground.

In the image rebinned to 0.1 ACIS pixels and smoothed
with a  0.492$\arcsec$ Gaussion filter,
the spatial distribution for \sdssj01 appears somewhat elongated (Figure\,\ref{fig1}).
To measure the extension of this source we use CIAO tool `srcextent'. While, due to the limited source counts of 14 to 15, no conclusive result is returned.
 We then manually extract the photometry of \sdssj01\  and found 14 counts within a 3 pixel ($\sim1.5\arcsec$) 
radius aperture, which corresponds to 94.7\% and 89.2\% of the encircled energy fraction in 0.5-2 keV and 2-7 keV, respectively\footnote{estimated with the CIAO tool SRC\_PSFFRAC}.
There is one more count detected at energy of 1.73 keV within the 4 pixel radius aperture (Figure\,\ref{fig1}).
According to the point-spread function at the source position, the expected count in Poisson statistics is 0.16 in the annulus with radii of 3 and 4 pixel, and 0.78 outside of the aperture with radius of 3 pixel.
That is, the chance of this photon belong to the quasar \sdssj01 is 14\%-36\%.
Thus for \sdssj01\ we restrict subsequent analysis to the detected photons within the 3 pixel radius aperture.
The average background level, 
estimated from an annulus in the source position with 
inner and outer radii of 15 and 20 arcsec, in an aperture of the source size is  0.06 in 0.5-2 keV and 0.05 counts in 2-7 keV.

For \sdssj01, the full-band (0.5-7 keV) X-ray counts are 13.9$_{-3.7}^{+4.8}$, 
with 11.9$_{-3.4}^{+4.6}$ in soft band (0.5-2 keV) and 2.0$_{-1.3}^{+2.6}$ in hard band (2-7 keV). 
The errors of the X-ray counts were computed according to \citet[][]{gehrels86}.
The hardness ratio, estimated with Bayesian method\footnote{http://hea-www.harvard.edu/astrostat/behr/} \citep[][]{park06}, is $-0.78_{-0.14}^{+0.07}$, indicating that the 
X-ray spectrum of \sdssj01\ is soft.
The inferred effective photon index from the hardness ratio is indeed large, at $3.2^{+1.0}_{-0.4}$, estimated  
with Xspec \citep[v.12.9;][]{arnaud96} and the latest calibration files.

We then perform basic spectral analysis for the quasar. 
The source spectrum, background spectrum, response matrix files (RMF) 
and auxiliary matrix files (ARF) are built using the CIAO script SPECEXTRACT.
The spectrum is grouped with a minimum of 2 counts per bin. 
Given that there is only $\sim 14$ net counts, we fitted the spectrum using XSPEC 
with a simple power law model and fixed Galactic absorption.
The C-statistic (cstat) was used due to the limited photon counts (Cash 1979).
The best fit photon index is $\Gamma$ = 3.03$^{+0.78}_{-0.70}$, with 
cstat = 8.31 for 5 degrees of freedom. 
We also perform Montecarlo simulations using the exposure time and response matrices, and fit the simulated one thousand spectra with the same model as above.
The value of inferred photon index is consistent with that from hardness ratio and best fit, with mean value of 3.13 and standard deviation of 0.71.

\begin{figure}
\includegraphics[width=0.35\textwidth, angle=-90]{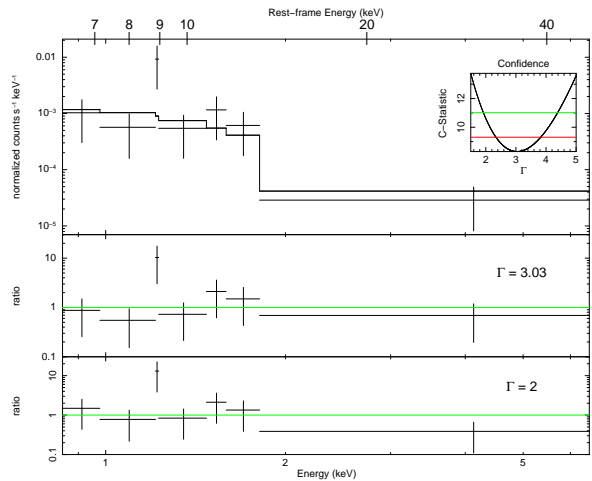}
\caption{Upper panel: \chandra\ spectrum of \sdssj01\ and the best-fit power-paw model with $\Gamma = 3.03$ ($N_{H}$ fixed at Galactic $N_{H}^{Gal}$). Inset, confidence curve for the fitted photon index. Middle: data to model ratio of the best-fit model. Significant residual at $\sim$ 1.2 keV (8.8 keV at rest-frame) is present in the fitting with different photon indices. 
Lower: data to model ratio, where a power-law with fixed $\Gamma=2.0$ is assumed.  
\label{fig2}}
\end{figure}

The best fit results and confidence curve for the photon index are shown in Figure\,\ref{fig2}. As a comparison, we also show the fitted spectrum with fixed photon index of 2.0, the typical value found for quasars. From the data to model ratios we can see that the model with soft spectral index of 3.03 improves the fitting at the most soft and hard energy bands. The rest-frame 2-10 keV luminosity implied by the best fit is 9.0$^{+9.1}_{-4.5}$ $\times$ 10$^{45}$ erg s$^{-1}$.
There is significant residual at $\sim$ 1.2 keV (8.8 keV at rest-frame). 
There is no known spectral feature at this energy in quasars, and more counts are needed to investigate the nature of this excess. 

\begin{figure}
\includegraphics[width=0.45\textwidth]{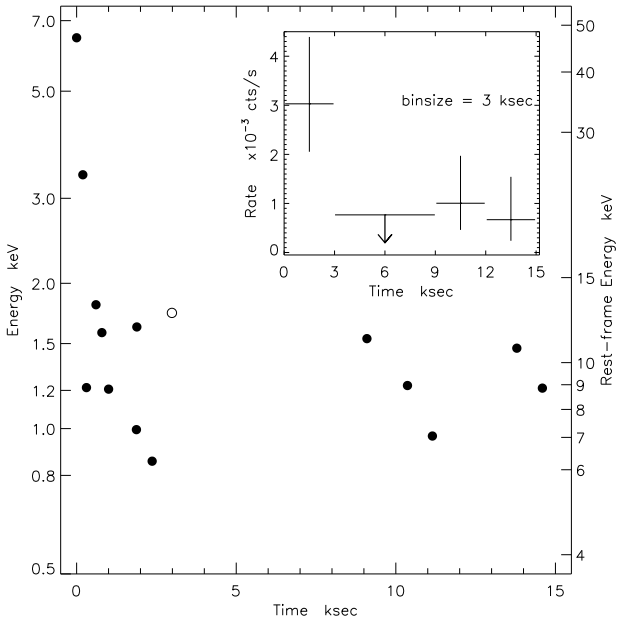}
\caption{Detected time vs. energy of each photon. The open circle labels the one more count detected within the 4 pixel radius around the source. Inset, light curve of \sdssj01\ in 0.5-7 keV with time bin size of 3 ks. None of the photon is detected in the middle 6 ks.  The 99\% confidence upper limit is shown.
\label{fig3}}
\end{figure}

We label the detected energy and receiving time of each photon in Figure\,\ref{fig3}. There are 4 photons detected between 1.207 keV and 1.228 keV. This unusual feature causes the significant residual at $\sim$1.2 keV in the spectral fitting (Figure 2).
Figure\,\ref{fig3} also shows the light curve of \sdssj01 with time bin size of 3 ks. There are 9 photons detected in the first 3 ks bin, while in the middle 6 ks interval no photon is detected, with 99\% confidence upper limit of 4.605 \citep[][]{gehrels86}. 
We first test the variation by analyzing the photon arrival time using the Gregory-Loredo algorithm \citep[][]{gregory92} with CIAO tool GLVARY.
The returned variability probability is P$_{\text{GL}}=$ 0.967, and the variability index is 6 from the test. 
We then test the chance for a random event of detecting 9 photons in one of the five bins for the total 14 photons with simulation. The chance is 186 in the 10$^{5}$ simulated light curves, corresponding to the probability of 0.019\%. Longer exposure observations are needed to investigate this possible variability. 

To investigate the broad band properties of \sdssj01 we calculate the X-ray-to-optical spectral energy distribution parameter, $\alpha_{\text{ox}}$, the slope of a nominal power law between 2500 \AA\ and 2 keV in the rest frame [\aox = 0.384$\times$log$(f_{2\text{ keV}}/f_{2500\text{ \AA}})$, where $f_{2 \text{ keV}}$ is 
the flux density at 2 keV and $f_{2500\text{ \AA}}$ is the flux density at 2500 \AA]. For \sdssj01, rest frame 2500\AA\ moves to the near-infrared at 18250 \AA\ in the observed frame, falling in the gap between H and K bands as shown in Figure 3 in \citet[][]{wu15}.  Using the ultraviolet-optical continuum slope measured in \citet[][]{wu15} we estimate the flux density at 2500 \AA\ to be {3.04}$\pm$ {0.45}$\times10^{-16}$ erg s$^{-1}$ cm$^{-2}$ \AA$^{-1}$. An energy of 2 keV in the rest frame of \sdssj01 corresponds to 0.273 keV in the observed frame. This energy is poorly sampled by ACIS. We compute the 0.273 keV flux and the uncertainty from the best fit power law model in XSPEC spectral fitting. This yields a value for \aox\ of $-1.22^{+0.07}_{-0.05}$, with errors dominated by the uncertainty of the rest-frame 2 keV flux.

\section{DISCUSSION}
The ultraluminous quasar \sdssj01\ is strongly detected in our exploratory \chandra\ observation with $\sim$14 counts,
We find that the X-ray properties of this quasar to be unusual in several aspects. 

First, there is a hint of extended emission, and this feature comes from photons in relatively soft energy bands. As first proposed in \citet[][]{fabian14}, relativistic electrons in powerful jetted objects with Lorentz factor $\Gamma\sim10^{3}$ can upscatter CMB photons into the soft X-ray band. Deeper XMM-Newton observation, which has much higher soft X-ray sensitivity compared to \chandra, would be ideal to investigate the possibility of this extended emission. 

Second, the X-ray spectrum of \sdssj01 is very soft with inferred photon index of 3.03$^{+0.78}_{-0.70}$, where the mean value of the X-ray photon index found in the samples of
low- and high-redshift quasars is 1.89 $< \langle\Gamma\rangle <$ 2.19 \citep[i.e.,][]{piconcelli05, just07,young09}. Although there are still debates about whether the X-ray spectrum of high-redshift quasars significantly differs from that of quasars at lower redshifts \citep[][]{just07,grupe06}, we note that the highest redshift quasar ULAS J1120+0641 may also exhibits a soft X-ray spectrum with photon index of $2.6^{+0.4}_{-0.3}$\citep[][see Moretti et al. 2014 for a different value of inferred $\Gamma$]{page14}. It has been suggested that the soft X-ray spectrum is an indication that the central massive black holes at early epochs are growing at high accretion rate \citep[][]{pounds95,leighly99}.

The inferred $\Gamma$ can be used as an estimator of  $L/L_{\text {Edd}}$ by using the linear correlation between $\Gamma$ and 
log($L/L_{\text {Edd}}$), 
where $L$ is bolometric luminosity and $L_{\text Edd}$ is Eddington luminosity \citep[][]{shemmer08, risaliti09, jin12, brightman13,yang15}. 
Using different relations presented in the literatures, we find that the corresponding $L/L_{\text {Edd}}$ for \sdssj01 has minimum value of 6 \citep[][]{risaliti09}, and can be significantly higher \citep[i.e.,][]{shemmer08}. If those relations did apply to \sdssj01, it could be accreting at significantly above the Eddington rate. Mildly super-Eddington accretion corresponds to a shorter e-folding timescale for black hole mass growth, which can ease the problem of assembling massive black holes out of stellar-mass seeds at early times even in the case of intermittent accretion \citep[][]{madau14}.

\begin{figure}
\includegraphics[width=0.5\textwidth]{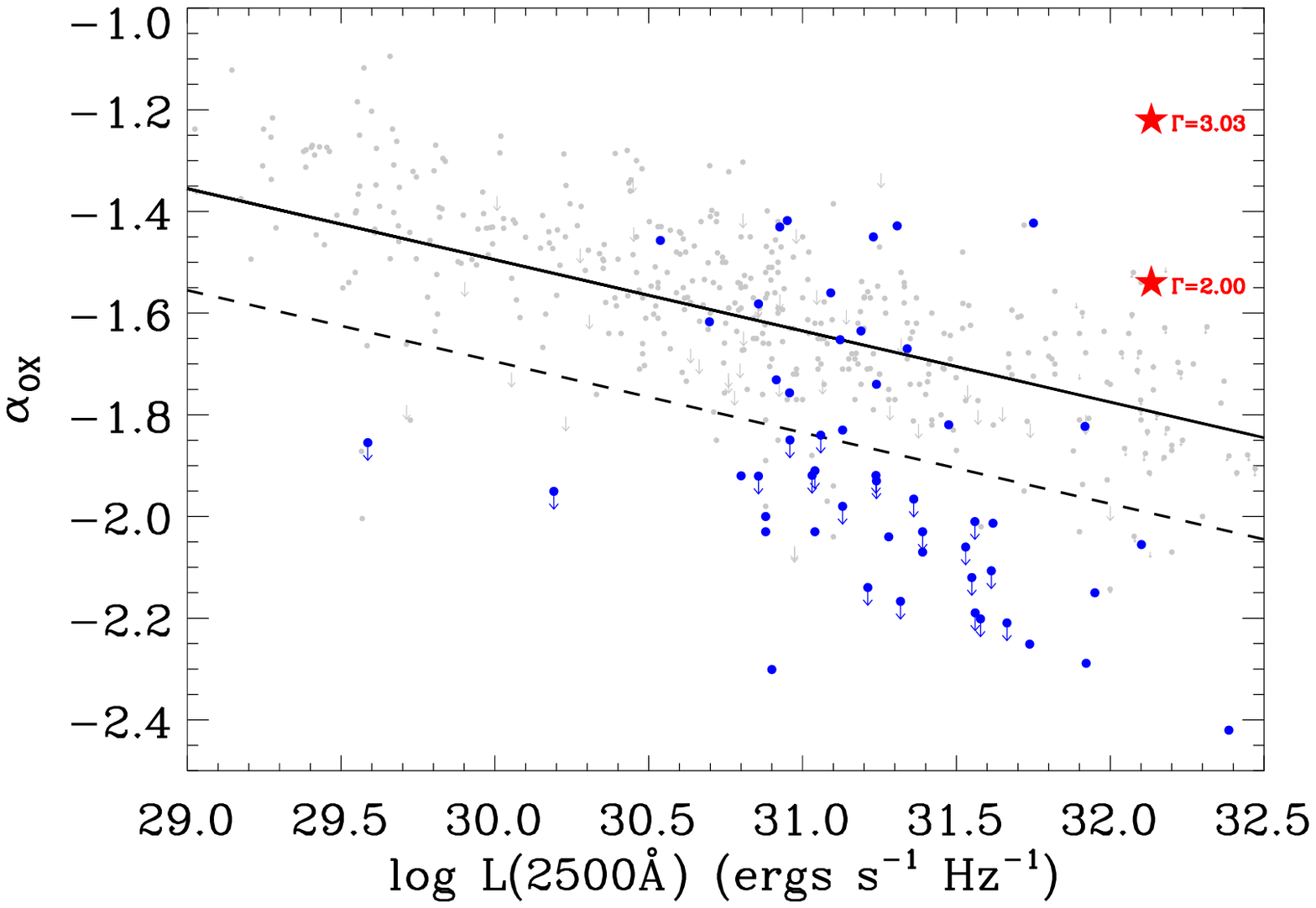}
\caption{Location of \sdssj01\ (red stars)  in the X-ray-to-optical power-law slope parameter \aox\  vs. 2500 \AA\ monochromatic luminosity. The grey dots are the quasars from the samples of Just et al. (2007), Steffen et al. (2006) and Gibson et al. (2008). The blue dots are the weak line quasars  and PHL 1811 analogs from \citet[][]{luo15}. The solid line represents the \citet[][]{just07} \aox-$L_{2500 \text \AA}$ relation, and the dashed line marks the devision between X-ray weak and X-ray normal quasars adopted in \citet[][]{luo15}. For \sdssj01 the two red stars represent the values of \aox\ derived with X-ray photon index of 3.03 and 2.0, respectively.
\label{fig4}}
\end{figure}

Third, \sdssj01\ has a relatively higher X-ray to optical flux ratio, compared to those found in other quasars of comparable ultraviolet luminosity (Figure\, \ref{fig4}). It is well established that the X-ray-to-optical power-law slope parameter \aox\ of quasars significantly correlate with the ultraviloet 2500 \AA\ monochromatic luminosity ($L_{2500 \text \AA}$, Steffen et al. 2006; Just et al. 2007). 
According to the \aox--$L_{2500 \text \AA}$ relation given by Just et al. (2007), a value of \aox\ $=-1.79\pm0.14$ is expected for \sdssj01,  compared to be the value of \aox\ = $-1.22^{+0.07}_{-0.05}$ obtained from \chandra\ observation.
The observed large \aox\ could be due to the inferred steep X-ray photon index, which is still not well constrained based on current S/N of the data.
While,  even if we adopt a flat X-ray spectral shape with the X-ray photon index of nominal 2.0 to infer the rest-frame 2 keV flux for \sdssj01, the derived \aox\ is $-1.54^{+0.07}_{-0.05}$,  is still larger than the average and is at upper envelop of the observed \aox\ values at the comparable ultraviolet luminosity (Figure\,\ref{fig4}). However, the value of \aox\ for \sdssj01\ is still in the range of the ones found in the quasars at low to median redshift, supporting the no evolution of \aox\ with redshift \citep[][]{brandt02, just07, risaliti15}

\sdssj01\ is a weak-line quasar with rest-frame equivalent width of the Ly$\alpha+\ce{Nv} \sim 10 \text\AA$ \citep[][]{wu15}. Weak-line quasars (WLQs) are a subclass of radio-quiet quasars that have almost extremely weak or undetectable emission lines \citep[e.g.][and references therein]{fan99,meusinger14}. The nature of WLQs is still under investigation,  although a number of hypotheses have been proposed, such as abnormal broad emission line region with significant deficit of line-emitting gas \citep[e.g.,][]{shemmer10} or in an early evolutionary stage of formation \citep[e.g.,][]{hryniewicz10}, a soft ionising spectral energy distribution due to intrinsic X-ray weakness \citep[e.g.,][]{leighly07} or due to small-scale absorption \citep[e.g.,][]{wu12,luo15}. Significant fractions ($\sim 50\%$) of the WLQs are distinctly X-ray weak compared to typical quasars \citep[][]{shemmer09,wu12,luo15}. Our results demonstrate that \sdssj01\ is definitely not an X-ray weak WLQ, as shown in Figure\,\ref{fig4}. If we interpret \sdssj01\ in the scenario proposed by Wu et al.(2012) and Luo et al.(2015), in which they unified X-ray weak and X-ray normal WLQs by orientation effect with proposed `shielding-gas', this quasar should be viewed at a small inclination angle.

Finally, we detect a hint of X-ray variability for \sdssj01\ within our exposure.
For high-redshift quasar at this luminosity, its X-ray flux is not expected to vary at the timescale of $\sim 10$ ks unless beaming is involved. 
Given that the light crossing time of even the Schwarzschild radius for a 12 billion $M_{\odot}$ black hole is much longer than 10 ks, the 
X-ray flux of \sdssj01 is not expected to vary in our observations unless strong beaming is involved. 
Deeper XMM-Newton observation will help to better constrain the spectral features, probe the extended emission and investigate the variability on both short ($\sim$1 hr) and longer (months to year) timescales for this ultraluminous quasar.

\acknowledgments
We thank Tinggui Wang, Andy Fabian, Rongfeng Shen and Yirong Yang for useful discussions, and B. Luo for providing the data of Figure 4. We also thank the anonymous referee for the helpful comments. F. W. and X-B. W. thank the support from the NSFC grants No.11373008 and 11533001, the Strategic Priority Research Program ``The Emergence of Cosmological Structures'' of the Chinese Academy of Sciences, grant No. XDB09000000, and the National Key Basic Research Program of China 2014CB845700.
This work is based observations made by the \chandra\ X-ray Observatory and has made use of software provided by the Chandra X-ray Center in the application package CIAO. Support for this work was provided by the National Aeronautics and Space Administration through Chandra Award Number GO5-16116B issued by the Chandra X-ray Observatory Center, which is operated by the Smithsonian Astrophysical Observatory for and on behalf of the National Aeronautics Space Administration under contract NAS8-03060

\clearpage

\end{document}